# Co-operative Influence of $O_2$ and $H_2O$ in the Degradation of Layered Black Arsenic

Mayank Tanwar[1+], Sagar Udyavara[1+*], Hwanhui Yun[1], Supriya Ghosh[1], K. Andre Mkhoyan[1*], Matthew Neurock[1,2*]

[1]*Department of Chemical Engineering and Materials Science, University of Minnesota, Minneapolis, Minnesota 55455, USA*

[2]*Department of Chemistry, University of Minnesota, Minneapolis, Minnesota 55455, USA*

[+]These authors contributed equally to the work

**\*Corresponding Authors**: udyav001@umn.edu (SU), mkhoyan@umn.edu (KAM), mneurock@umn.edu (MN)

**Abstract**

Layered black arsenic (b-As) has recently emerged as a new anisotropic two-dimensional (2D) semiconducting material with applications in electronic devices. Understanding factors affecting the ambient stability of this material remains crucial for its applications. Herein we use first-principles density functional theory (DFT) calculations to examine the stability of the (010) and (101) surfaces of b-As in the presence of oxygen ($O_2$) and water ($H_2O$). We show that the (101) surface of b-As can easily oxidize in presence of $O_2$. In the presence of moisture contained in air, the oxidized b-As surfaces favorably react with $H_2O$ molecules to volatilize As in the form of $As(OH)_3$ and $AsO(OH)$, which results in the degradation of the b-As surface, predominantly across the (101) surface. These predictions are in good agreement with experimental electron microscopy observations, thus demonstrating the co-operative reactivity of $O_2$ and $H_2O$ in the degradation of layered b-As under ambient conditions.

The examination of two-dimensional (2D) materials has led to many scientific advances and ground breaking applications in the recent past.[1–4] Among the various studied 2D materials, layered black arsenic (b-As) is emerging as a new elemental 2D material from the pnictogen group with characteristic properties comparable to or better than those of black phosphorus (BP).[5–7] Despite being synthesized only recently,[5,6,8,9] b-As has already shown desirable layer-number-dependent electronic properties and the potential for being one of the building block materials for next-generation opto-electronic devices.[6,10,11] A more detailed understanding of black arsenic's ambient stability and establishing ways to improve it, however, are required for its utilization. The study by Yun et al.[12] showed that b-As degrades at atmospheric conditions, and the rate of degradation is sensitive to the humidity of the air. It was observed that, despite structural similarities with BP, b-As degrades primarily by volatilization of As atoms from the surface. Yun et al.[12] also showed that b-As flakes, when kept at elevated humidity, experience strong directional etching along particular crystallographic planes that are perpendicular to its layers. The volatilization of As atoms from b-As structure suggests that during ambient degradation, As likely reacts with $O_2$ and $H_2O$ molecules present in the surrounding atmosphere to form volatile arsenic species. The observed degradation of b-As is in stark difference with ambient degradation of BP, which is driven mainly by the formation of amorphous $P_xO_y$.[13–22] Thus, to close the gap in understanding the chemical transformations that lead to b-As degradation at ambient conditions, in this work, we examine in detail the interaction of $O_2$ and $H_2O$ molecules with the different facets of the b-As and their role in its degradation using first-principles density functional theory (DFT) calculations. We propose a mechanism for surface As removal and degradation of b-As under ambient conditions. The mechanism involves an energetically favorable sequential reaction sequence where $O_2$ first reacts with As to partially oxidize the As surface. Water then subsequently reacts with the oxidized As surface to form volatile $As(OH)_3$ and $AsO(OH)$ intermediates. The key predictions from this theoretical study are validated by comparing with results from scanning transmission electron microscopy (STEM) measurements, which are also presented and discussed herein.

**Results and Discussion**

STEM studies of b-As ambient stability show that when thin b-As films are kept under regular ambient conditions or in humid air, it disintegrates (Fig. 1), and this disintegration is facilitated when the relative humidity of the air is high. The directional etching along specific crystalline planes, perpendicular to layers, appears to be the dominate process (Fig. 1b). While etching along As {201}, {101}, and {102} planes was observed, etching along the {101} plane was the most common.[12] These observations also suggest that when b-As reacts with $H_2O$ and $O_2$ molecules present in the air, it will result in the formation of As-containing gaseous compounds.

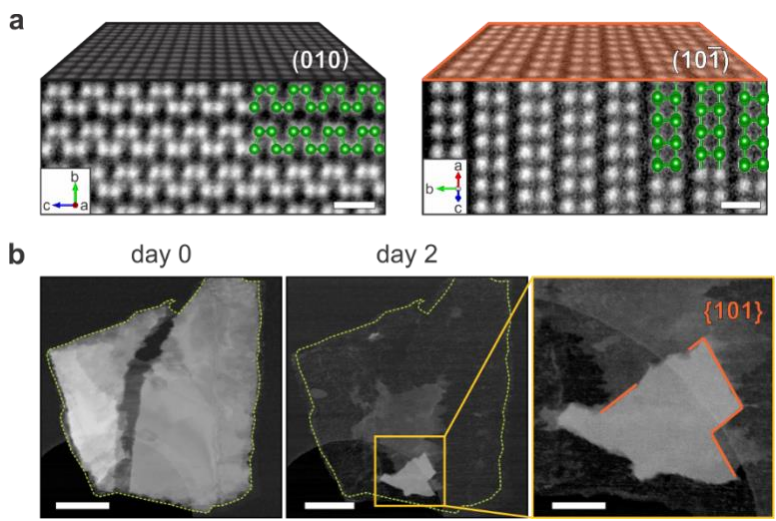

**Figure 1.** High-angle annular dark-field (HAADF)-STEM images of exfoliated b-As. (a) Structures of the b-As constructed using atomic-resolution HAADF-STEM images viewed along the [010] and [101] crystallographic directions. Van der Waal gaps between layers are visible in both directions. Atomic models are overlaid on the side of the images. Scale bars are 0.4 nm. (b) HAADF-STEM images, taken along the perpendicular to b-As layers (*b*-direction), showing the degradation of a b-As flake: as-prepared (day 0) and after being stored in 98% humid air for two days. The initial shape of the flake is outlined with a dashed line. A small and mostly intact section of the b-As flake shows strong directional etching along {101} crystallographic planes. Scale bars are 400 nm.

To evaluate the reactivity of water and oxygen with b-As, models of the b-As films with different exposed surfaces were constructed using the atomic structure of the bulk b-As crystal.[23] These models were then optimized to their establish their lowest energy configurations using *ab initio* DFT simulations (for details, see SM Fig. S1). These optimized surfaces were subsequently used to determine the adsorption and dissociation energies of water and $O_2$. Two surface terminations of the b-As crystal that were examined in depth were: the (010) surface, which is parallel to the van der Waals bonded layers, and the easy-etch (101) surface, which is one of the side surfaces perpendicular to layered (010) surface. It should be noted that in the case of the exposed (101) surface, the presence of dangling covalent bonds resulted in slight changes in the atomic configuration of the surface upon optimization with DFT (for details, see SM, Fig. S1).

To evaluate the adsorption of water on b-As surfaces, $H_2O$ molecules were initially placed at low coverages of $1/8^{th}$ of a surface on the (010) surface and $1/16^{th}$ of a surface on the (101) surface (for details, see SM, Fig. S2), and subsequently optimized to determine their lowest energy configurations. After optimization, the initially bound $H_2O$ molecule tends to move further away from the surfaces with a final As-O distance of 3.23 Å for the (010) and 3.28 Å for the (101) surface (Fig. 2) thus resulting in adsorption energies on the (010) and (101) surfaces of $\Delta E_{ads}$ = -16 and -19 kJ/mol, respectively. These slightly exothermic energies, along with large interatomic distances, indicate that water is weakly bound to the surface As atoms through weak long-range interactions. Therefore, the adsorption of water to the surfaces of thin b-As films kept under ambient, or even humid conditions, is not favored, and, as such, the covalent-type interactions between the water molecules and surface As atoms are unlikely.

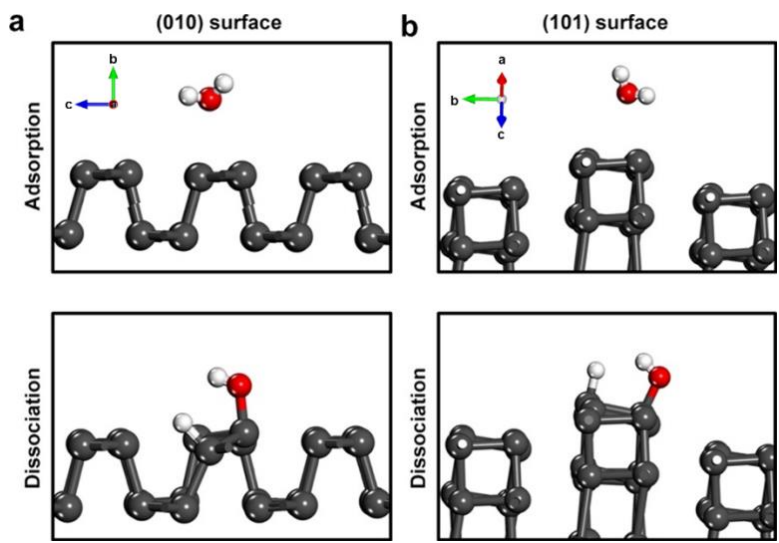

**Figure 2.** DFT-optimized structures for the adsorption (top) and dissociation (bottom) of $H_2O$ on (a) (010) and (b) (101) surfaces of b-As. In these structures, As atoms are in dark grey color, O in red, and H in white. The adsorption energies and interatomic distances for these structures are reported in Table 1.

The dissociation of a H$_2$O molecule over two vicinal As atoms on both b-As (010) and (101) surfaces was also considered. Water was found to dissociate over these surfaces resulting in the formation of As-H and As-O(H) bonds with distances of 1.54 and 1.89 Å, respectively, for the (010) surface, and 1.57 and 1.86 Å, respectively for the (101) surface (Fig. 2). The dissociation of H$_2$O leads to the formation of strong covalent bonds with the surface As atoms (see Table 1), which tend to pull the As atoms away from the surfaces as shown in Fig. 2. The energy for the dissociative adsorption of the H$_2$O molecule on the b-As surfaces, however, was found to be highly endothermic with dissociation energies as high as $\Delta E_{diss}$ = 163 and 101 kJ/mol for (010) and (101) surfaces, respectively, thus making dissociation very unlikely to occur. Therefore, it can be concluded that water alone will not likely react with b-As to cause the experimentally observed degradation.

The role of oxygen in the degradation of b-As was subsequently evaluated by considering the adsorption of O$_2$ on the (010) and (101) surfaces. Three different modes of adsorption of O$_2$ over the b-As surfaces were examined: (i) molecular O$_2$ adsorption via di-σ bonding where the two oxygen atoms of O$_2$ form covalent bonds to two vicinal surface As atoms; (ii) molecular O$_2$ adsorption via π-bonding where O$_2$ binds to a single As atom via a single π-bond; and (iii) dissociative O$_2$ adsorption over two vicinal surface As atoms. The optimized structures for the dissociative adsorption mode are shown in Fig. 3 and the resulting As-O and O-O distances along with adsorption and dissociative reaction energies are summarized in Table 1(see SM Fig. S3 for di-σ and π-bonding modes of adsorption). The results indicate that only the dissociative O$_2$ adsorption is exothermic on both surfaces, while the other two modes of molecular absorption are endothermic. It should be noted that, while the gas phase O$_2$ is in a triplet spin state with two unpaired electrons, the dissociative adsorption of O$_2$ leads to the formation of two singlet oxygen atoms that covalently bind to As atoms on the surface. The strong covalent bonds that form between the surface As and the O atoms result in the high exothermic reaction energies of $\Delta E_{diss}$ = -163 and -164 kJ/mol for the (010) and the (101) surfaces, respectively. Further evaluation of the transition states and activation energies for O$_2$ dissociation show a higher barrier of 76 kJ/mol for dissociation over the (010) surface and a much lower barrier of 17 kJ/mol for dissociation over the (101) surface (Fig. S5(b)). This can be attributed to dissociation over the saturated rigid framework of the (010) surface as compared to the reconstructed (101) surface which initially had unsaturated dangling bonds prior to reconstruction. The low barriers observed over the (101) surface compared to (010) surface suggests that the (101) surfaces will be more easily oxidized compared to the relatively more stable (010) surface which, as we shall see in the succeeding sections, leads to preferential degradation along the (101) plane. Thus, it can be speculated that if the b-As films are kept at ambient conditions with oxygen in the air, the exposed surfaces will begin to oxidize predominantly via $O_2$ dissociation over the (101) surface.

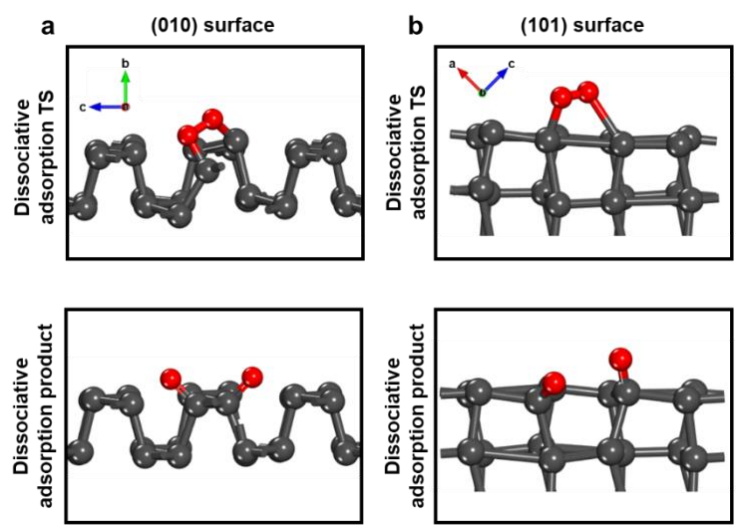

**Figure 3.** DFT-optimized transition (top) and product state (bottom) structures for dissociative $O_2$ adsorption on (a) (010) and (b) (101) surfaces of b-As. The dissociative reaction energies and activation barriers along with the reaction energies for the other adsorption modes and their respective interatomic distances are listed in Table 1. In these structures, the As atoms are shown in dark grey color, O in red, and H in white.

**Table 1.** Interatomic distances and the adsorption and dissociative reaction energies for the different $O_2$/$H_2O$ adsorption and reaction steps that can occur over the (010) and (101) surface terminations of b-As.

| Reaction step | Bond | Bond distances (Å) | | Adsorption and dissociative reaction energies (kJ/mol) | |
|---|---|---|---|---|---|
| | | (010) | (101) | (010) | (101) |
| $H_2O$ adsorption | As-O | 3.234 | 3.283 | -16 | -19 |
| $H_2O$ dissociation | $As_1$-H | 1.540 | 1.565 | 163 | 101 |
| | $As_2$-O | 1.898 | 1.864 | | |
| $O_2$ di-σ adsorption | $As_1$-$O_1$ | 2.218 | 1.952 | 63 | 2 |
| | $As_2$-$O_2$ | 2.225 | 2.024 | | |
| | $O_1$-$O_2$ | 1.350 | 1.446 | | |
| $O_2$ π-adsorption | As-$O_1$ | 1.914 | 1.908 | 51 | 20 |
| | As-$O_2$ | 1.832 | 1.826 | | |
| | $O_1$-$O_2$ | 1.573 | 1.560 | | |
| $O_2$ dissociative adsorption | $As_1$-$O_1$ | 1.692 | 1.675 | -163 | -164 |
| | $As_2$-$O_2$ | 1.692 | 1.669 | | |
| $O_2$ dissociative adsorption TS | $As_1$-$O_1$ | 1.871 | 2.041 | 76 | 17 |
| | $As_2$-$O_2$ | 1.977 | 2.804 | | |
| | $O_1$-$O_2$ | 1.503 | 1.337 | | |

The dissociative adsorption of oxygen results in the formation of the more active b-A sites on the As surface that may work together to drive the heterolytic dissociation of water. We therefore examined the adsorption of water and its dissociation on the partially oxidized b-As surfaces. The adsorption of $H_2O$ on the partially oxidized As sites was examined on the b-As surfaces with dissociated $O_2$ (1/4$^{th}$ of a surface O* coverage on the (010) surface and 1/8$^{th}$ of a surface O* coverage on the (101) surface). Interestingly, similar to the molecular $H_2O$ adsorption on a bare As surface, the optimized structures showed no tendency for $H_2O$ to molecularly adsorb on the oxidized surface as water moves away from the surface (see SM, Fig. S4).

However, when one considers the dissociative adsorption of $H_2O$ onto the partially oxidized b-As surface (for details, see SM discussion and Fig. S5), the adsorption becomes more favorable where the controlling reaction energies depend on the surface termination (Fig. 4). The dissociation of $H_2O$ was calculated to be slightly endothermic for the (010) surface of b-As with a relatively higher barrier ($\Delta E_{diss}$ = 5 kJ/mol, $\Delta E^{\ddagger}$ = 12 kJ/mol) whereas the dissociation over the (101) surface was calculated to be exothermic with an overall dissociation energy of $\Delta E_{diss}$ = -125 kJ/mol and no activation barrier. The water binds and heterolytically dissociates over two As(O)-As(O) vicinal sites on the surface to form an As(OH)-As(O)(OH) surface intermediate as shown in Fig. 4. Interestingly, the barriers to dissociate water over a single As(O) site are higher than over two As(O)-As(O) vicinal sites for both surfaces (for details, see SM, discussion and Figs. S5).

The adsorption and dissociation of a second $H_2O$ molecule at the As(OH)-As(O)(OH) site to form an As(OH)$_3$ species was calculated to be highly exothermic on both surfaces: $\Delta E_{diss}$ = -122 kJ/mol for the (010) surface, and $\Delta E_{diss}$ = -70 kJ/mol for the (101) surface, but proceeded with a higher barrier of $\Delta E^{\ddagger}$ = 54 kJ/mol for the (010) surface, and a relatively lower barrier of $\Delta E^{\ddagger}$ = 35 kJ/mol for the (101) surface. The As(OH)$_3$ species that forms is stable and ultimately desorbs from the b-As surfaces leaving behind a vacant site (Fig. 4). This 3-step reaction process, which first includes the dissociation of $O_2$ and subsequent dissociation of two $H_2O$ molecules on two vicinal As surface sites, is both kinetically as well as thermodynamically favorable over the (101) surface as compared to the (010) surface and thus likely leads to degradation of the b-As films observed in STEM experiments (Fig. 1).

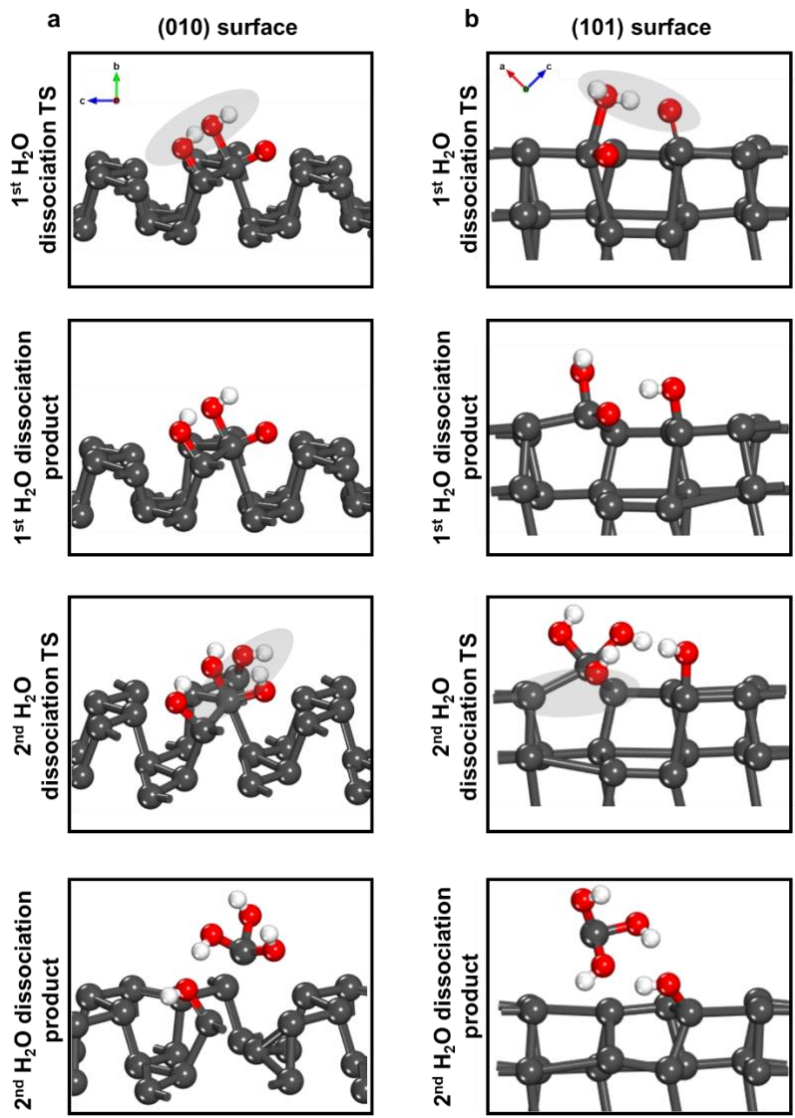

**Figure 4.** DFT-optimized transition and product state structures for dissociation of the 1st and 2nd H$_2$O molecule at the oxidized (a) (010) and (b) (101) b-As surfaces resulting in volatilization of As atom as As(OH)$_3$ molecule. The highlighted grey regions corresponds to bond making and breaking in the transition structures. In these structures, As atoms are in dark grey color, O in red, and H in white.

The formation of As(OH)$_3$ and its removal from the b-As surfaces creates a vacancy on the film surface. Since each As is coordinated to three other As atoms, this vacancy gives rise to three unsaturated As atoms. In addition, vacancies on the b-As surface are also expected to be present due to the finite ambient temperature and surface distortions, which may affect the stability of the surface. As such, we also evaluated the feasibility for the adsorption and dissociation of $O_2$ and $H_2O$ molecules around a vacancy site over both (010) and (101) b-As surfaces. Since each As atom is initially covalently bonded to three other As atoms, three dangling unsaturated As atoms are present at the vacancy. The relaxation of these resulting surfaces resulted in the reconstruction of the (101) surface where two of these unsaturated As atoms recombine, still leaving one unsaturated As atom as shown in Fig. 5 (for details, see SM Fig. S6). No such surface reconstruction was observed for the (010) surface, leaving all three As atoms uncoordinated.

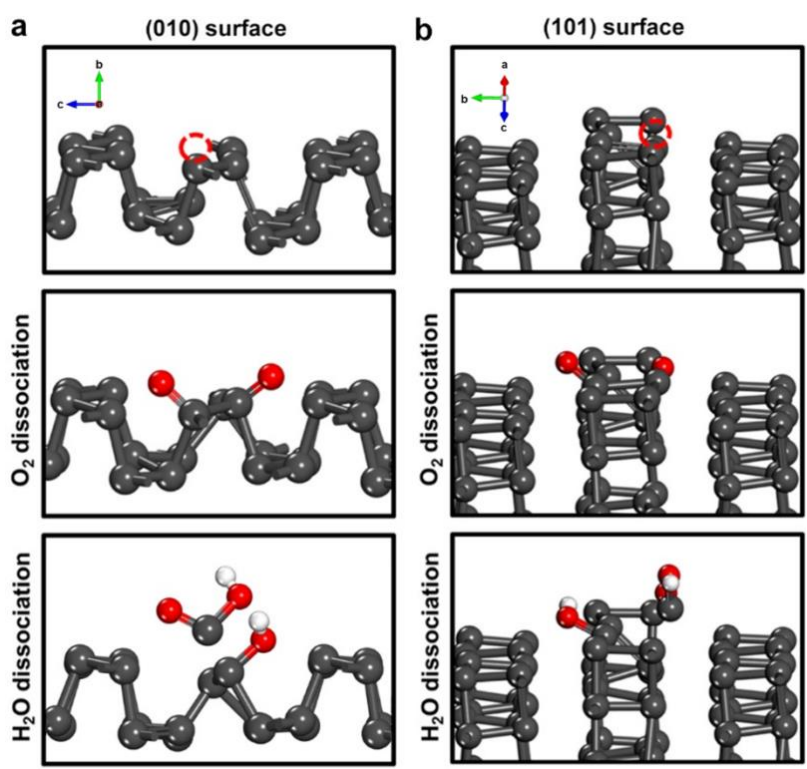

**Figure 5:** DFT-optimized structures for the (010) and (101) b-As surfaces showing a vacancy on the surface (top, the missing site is highlighted with a red dotted circle), dissociation of $O_2$ (center) and the first water molecule (bottom) at the As sites adjacent to the vacancy on the (a) (010) and (b) (101) b-As surfaces resulting in volatilization of an As atom as AsO(OH) molecule. In these structures, As atoms are in dark grey color, O in red, and H in white.

The computational results for the adsorption and dissociation of water at an As site adjacent to the vacancy showed that adsorption and dissociation at this site was unlikely. This is similar to results on the pristine As surfaces (for details see SM Fig. S7). On the other hand, as with the defect-free As surfaces, the dissociative adsorption of $O_2$ was calculated to be very favorable with highly exothermic adsorption energies of $\Delta E_{diss}$ = -242 kJ/mol for (010) surface and $\Delta E_{diss}$ = -227 kJ/mol for (101) surface (Fig. 5) (see also SM Fig. S8). The molecular adsorption of water at these oxidized As sites adjacent to the vacancy was again found to be unfavorable (see SM Fig. S9). The dissociation of water at these oxidized As site adjacent to the vacancy, on the other hand, was calculated to be favorable with dissociation energies of $\Delta E_{diss}$ = -57 kJ/mol for (010) surface and $\Delta E_{diss}$ = -100 kJ/mol for (101) surface (Fig. 5) (see also SM Fig. S10). Here, an As atom is removed from the surface as AsO(OH) (arsenious acid) that forms upon the dissociation of the first $H_2O$ as shown in Fig. 5. Hence, once the vacancy forms, the disintegration of that surface becomes even more facile. These results thus suggest that the degradation of b-As should be further accelerated in presence of these vacancies.

Fig. 6 summarizes the energy profile for the overall degradation mechanism over the pristine and defective (010) and (101) surfaces. As can be seen from the profile, the initial oxygen dissociation step has the highest activation barrier for the (010) surface and hence is postulated to be limiting the rate of degradation over this surface. While the dissociative adsorption of oxygen is thermodynamically facile over both the (101) and the (010) surfaces, this step is kinetically favored over the (101) surface with a relatively low barrier of 17 kJ/mol compared to 76 kJ/mol over the (010) surface. The subsequent dissociation of the first water molecule is also energetically favorable over the (101) surface with an exothermic reaction energy of -125 kJ/mol and no activation barrier. This is in contrast with the results for the (010) surface, where the dissociation of the first water molecule is slightly endothermic with a dissociation energy of 5 kJ/mol and activation barrier of 12 kJ/mol. While the $2^{nd}$ water dissociation is more exothermic over the (010) surface ($\Delta E_{rxn}$ = -122 kJ/mol) compared to that over the (101) surface ($\Delta E_{rxn}$ = -70 kJ/mol), due to the favorability for the initial oxygen dissociation and water dissociation steps, the reaction energy for the overall cycle leading to As removal is significantly more exothermic for the (101) surface ($\Delta E_{rxn}$ = -359 kJ/mol) as compared to the (010) surface ($\Delta E_{rxn}$ = -280 kJ/mol). This suggests that the subsequent degradation is expected to predominantly proceed across the (101) surface, thus leading to the formation of As-vacancies. The formation of the As-vacancy sites in turn results in the coordinative unsaturation of the As atoms that are more reactive thus leading to higher exothermic dissociation energies, as shown in Fig. 6, and more facile and faster degradation of the b-As surface. The degradation likely occurs via a layer-by-layer peeling of the As surface wherein the removal of the first As atom is rate-determining. Once the first As atom is removed and a vacancy site is created, the degradation process becomes energetically more favorable and the whole layer can then rapidly disintegrate.

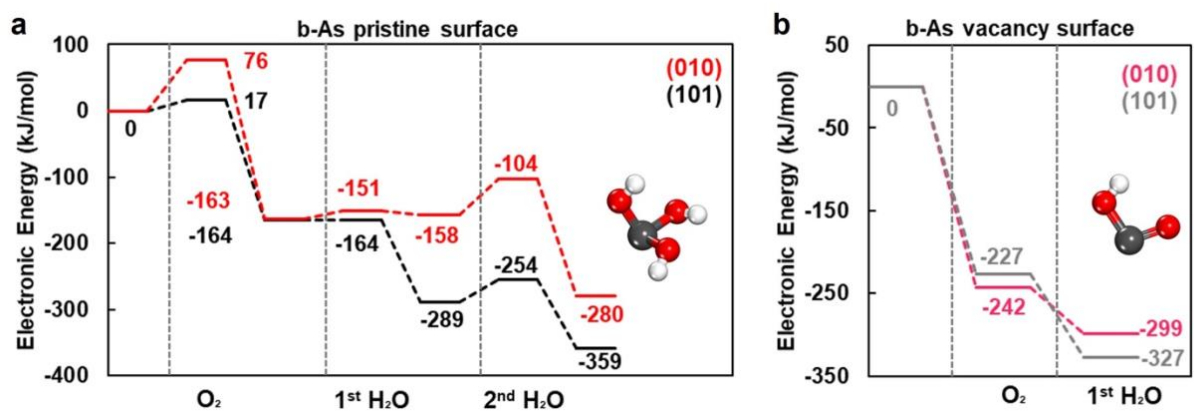

**Figure 6.** The reaction energy profile comparing the reaction barriers and energies for the dissociation of $O_2$ and co-dissociation of the first and second $H_2O$ molecules over the (010) surface (red) and the (101) surface (black) of b-As. (a) The 3-step-reaction on the pristine b-As surfaces results in formation of volatile $As(OH)_3$. The reaction energy profile comparing the reaction energies for the dissociation of $O_2$ and co-dissociation of the first $H_2O$ molecule over the (010) vacancy surface (pink) and the (101) vacancy surface (grey) of b-As. (b) The 2-step-reaction at the surface vacancy site results in formation of volatile $AsO(OH)$.

The predictions from the DFT calculations agree rather well with the experimental observations. When b-As flakes were kept in a dry air environment, with only 1-5 % relative humidity, they only become slightly oxidized without visible changes in flake structure, as shown in Fig. 7(a). This is consistent with the prediction of a favorable exothermic $O_2$ adsorption on the b-As surfaces with minimal effects on the integrity of the b-As atomic structure. Then, when the b-As flake is exposed to high humidity (98% relative humidity), directional etching across the {101} planes take place also in agreement with the computational results that show the oxygen and water dissociation steps are more facile over the {101} surfaces compared to that over the {010} surfaces. When b-As is placed in ambient air, under not a particularly high H₂O level, the role of vacancies on the surface degradation becomes important. As seen in Fig. 7(b), b-As flake with defects and vacancies shows that those regions around the surface As-vacancies tend to degrade faster as predicted.

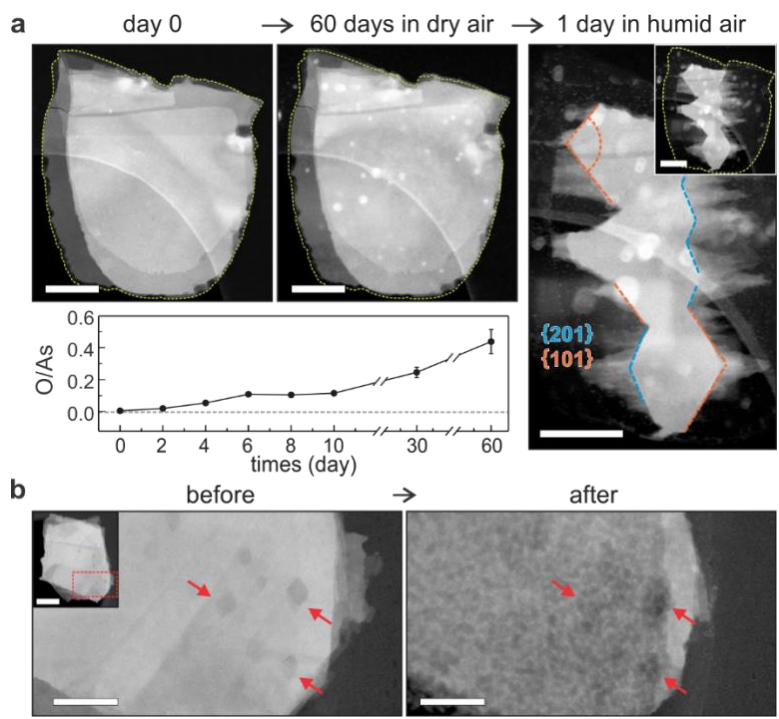

**Figure 7.** (a) HAADF-STEM images showing the degradation of a b-As flake under different environmental conditions: as-prepared b-As flake (day 0); the b-As flake after stored in dry air for 60 days and after keeping in humid air for one day. The initial shape of the flake is outlined with a dashed line. STEM-EDX compositional analysis shows that when flake is kept in dry air, it accumulates oxygen on its surfaces. All scale bars are 200 nm. (b) HAADF-STEM images of b-As flake showing morphology changes before and after being exposed to ambient air for < 1 hour. Scale bars are 100 nm. Inset image shows the shape of the flake, and the square in the inset indicates the magnified region. Scale bar in the inset is 300 nm. The regions with initial surface damage (indicated with red arrows) have more extended degradation.

In conclusion, while H₂O doesn't seem to react with b-As, in the presence of O₂, it can dissociate and disintegrate b-As. DFT calculations suggest that a plausible mechanism that can result in the degradation of pristine b-As involves three sequential dissociative adsorption steps. The first step involves the rate limiting dissociation of molecular oxygen over vicinal As atoms on the surfaces. A water molecule can then dissociate heterolytically over the As(O)-As(O) surface site to form a surface As(OH)-As(O)(OH) intermediate. A second water molecule can then dissociate heterolytically over this intermediate surface site to form the As(OH)$_3$ species which subsequently desorbs into the gas phase. The overall degradation process was calculated to be more exothermic for the (101) surface than the (010) surface. In addition, the high activation barrier for the initial oxygen dissociation over the (010) surface limits the initial oxidation and hence the onset of surface degradation. These differences should result in a slow degradation of b-As along the (010) surface compared to the (101) surface. Further, the formation of As-vacancies leads to more favorable and thus faster degradation of the surface via the formation of AsO(OH) species in a two-step-reaction requiring only a single water dissociation. All these predictions are consistent with experimental STEM observations, including that, in absence of water in the air, b-As undergoes surface oxidation with minimal effects on the integrity of the b-As surface structure. The results presented herein describe the critical co-operative role of oxygen and water in degradation of the b-As. It also suggests that keeping the b-As film in an environment where water and O$_2$ are not present simultaneously will minimize the degradation and dramatically prolong its stability.

**Methods**

*DFT Calculations.* All of the calculations described herein were carried out using plane-wave periodic density functional theory (DFT) methods, as implemented in the Vienna *Ab initio* simulation program (VASP).[24–27] The Revised Perdew-Burke-Ernzerhof (RPBE) functional form of the generalized gradient approximation was used along with the Pade Approximation for all of the calculations discussed to determine the corrections to the energy due to exchange and correlation effects.[28] PAW pseudopotentials were used to describe the interactions between the valence and core electrons.[29] D3 dispersion corrections, as proposed by Grimme, were used to account for the long-range interactions present in the system under study.[30] The structures used in the calculations were constructed from the bulk orthorhombic As crystal with lattice constants of a = 3.71 Å, a = 11.44 Å, and c = 4.69 Å.[23] For the (010) surface, the bulk As crystal was cleaved across the (010) plane, and a 2x2 supercell consisting of six layers (or three monolayers) was constructed. An eight-layered (101) surface was similarly constructed with a supercell size of 2×2. A vacuum slab of size 10 Å was inserted between each of these surfaces to avoid spurious interactions across the periodic unit cell along the z-direction (c-axis). The bottom 2 layers of the slab were fixed to their bulk calculated values in all the calculations for both the surfaces. Gamma centered k-point meshes of 3×3×1 and 4×2×1 were used to sample the Brillouin zone (BZ) for the (010) and (101) surface supercells, respectively. The plane waves used in the DFT calculations were constructed using an energy cut-off value of 400 eV. The tolerance for the energy convergence in the SCF cycle was set to $10^{-6}$ eV. The geometric relaxation was carried out until a maximum force of 0.05 eV/atom was reached.

Transition states were determined, herein, by using a two-step approach. The nudged elastic band (NEB) method[31,32] is used in the first step to establish the location of the transitions. The transition state location is subsequently refined in the second step using the dimer method.[33] In the NEB method, a reaction path is initially described by interpolating a series of images along the reaction coordinate joining reactants to products, which are then optimized until the tangential and normal forces were less than 0.08 eV/Å. The highest energy structure along the reaction coordinate was then used as an initial guess for the dimer method to converge the forces to less than 0.05 eV/Å and isolate the transition state. Spin polarized calculations using ISPIN = 2 tag were performed in the NEB and DIMER calculations for $O_2$ adsorption and dissociation to account for the triplet nature of $O_2$ in the gas phase which then forms singlet oxygen atoms on the surface upon dissociation.

***STEM Measurements.*** STEM experiments were performed using aberration-corrected FEI Titan G2 60-300 (S)TEM operated at 200 keV beam energy. HAADF-STEM imaging was carried out using a beam current of ~30 pA and probe convergence angle of 17.2 mrad. ADF detector inner and outer angles for HAADF-STEM imaging were 55 and 200 mrad, correspondingly. b-As flakes used in this study were prepared from b-As crystal purchased from 2D Semiconductors Inc. Plan-view TEM samples were prepared by mechanically exfoliating bulk b-As. The detailed description of TEM sample preparation is reported in Yun et al.[12] The humid air-conditioned and dry air-conditioned samples were stored under an ambient pressure of 760 torr and temperature of 20 ± 0.2 °C. The humidity level was controlled by locating deionized water and desiccants (Calcium sulfate from Sigma-Aldrich) alongside each sample in a closed-glass chamber for humid and dry conditions, respectively. Humidity and temperature of the closed-glass chamber were continuously measured with relative humidity maintained at 96–100 % for humid and 0.6–5 % for dry conditions.

**Acknowledgements**

This project was partially supported by NSF Center for Synthetic Organic Electrochemistry (CHE-2002158), UMN MRSEC programs (DMR-201140), and SMART, one of seven centers of nCORE, a Semiconductor Research Corporation program, sponsored by NIST. Parts of this work were carried out in UMN Characterization Facility supported in part by the NSF through the UMN MRSEC program. We also thank the Minnesota Supercomputing Institute (MSI) for use of its computational resources. We also would like to thank Prof. S. Koester and P. Golani for providing b-As samples.

**Associated content**

*Author Contributions*

SU, KAM, and MN conceived and designed the project. MT performed the DFT calculations with inputs from SU and MN. HY and SG carried out sample preparation, and HY performed STEM experiments and data analysis with inputs from KAM. MT, SU, KAM, and MN analyzed the results of calculations and wrote the manuscript with contributions from all authors.

*Supporting Information Available*

Optimization of (010) and (101) b-As surfaces, guess structures for molecular water adsorption on the (010) and (101) b-As surfaces, di-σ adsorption and π-adsorption of $O_2$ on (010) and (101) b-As surface, molecular water adsorption on the oxidized (010) and (101) b-As surfaces, discussion on configurational sampling for the dissociation of water over the oxidized As surfaces, flow diagrams of possible mechanistic routes for oxygen and two water molecules dissociation over the (010) and (101) b-As surfaces, optimization of (010) and (101) b-As surfaces with As vacancy, adsorption and dissociation of the $H_2O$ molecule on (010) and (101) b-As surfaces with As vacancy, di-σ adsorption and π-adsorption of $O_2$ and $H_2O$ on (010) and (101) b-As surface with As vacancy. This material is available free of charge via the Internet at http://pubs.acs.org.

Supplementary Material

# Co-operative Influence of $O_2$ and $H_2O$ in the Degradation of Layered Black Arsenic

Mayank Tanwar[1+], Sagar Udyavara[1+*], Hwanhui Yun[1], Supriya Ghosh[1], K. Andre Mkhoyan[1*], Matthew Neurock[1,2*]

[1]*Department of Chemical Engineering and Materials Science, University of Minnesota, Minneapolis, Minnesota 55455, USA*

[2]*Department of Chemistry, University of Minnesota, Minneapolis, Minnesota 55455, USA*

[+]These authors contributed equally to the work

***Corresponding Authors**: udyav001@umn.edu (SU), mkhoyan@umn.edu (KAM), mneurock@umn.edu (MN)

**Table of Contents**

Supplementary Figures S1 to S10

Supplementary Discussion

**Supplementary Figures**

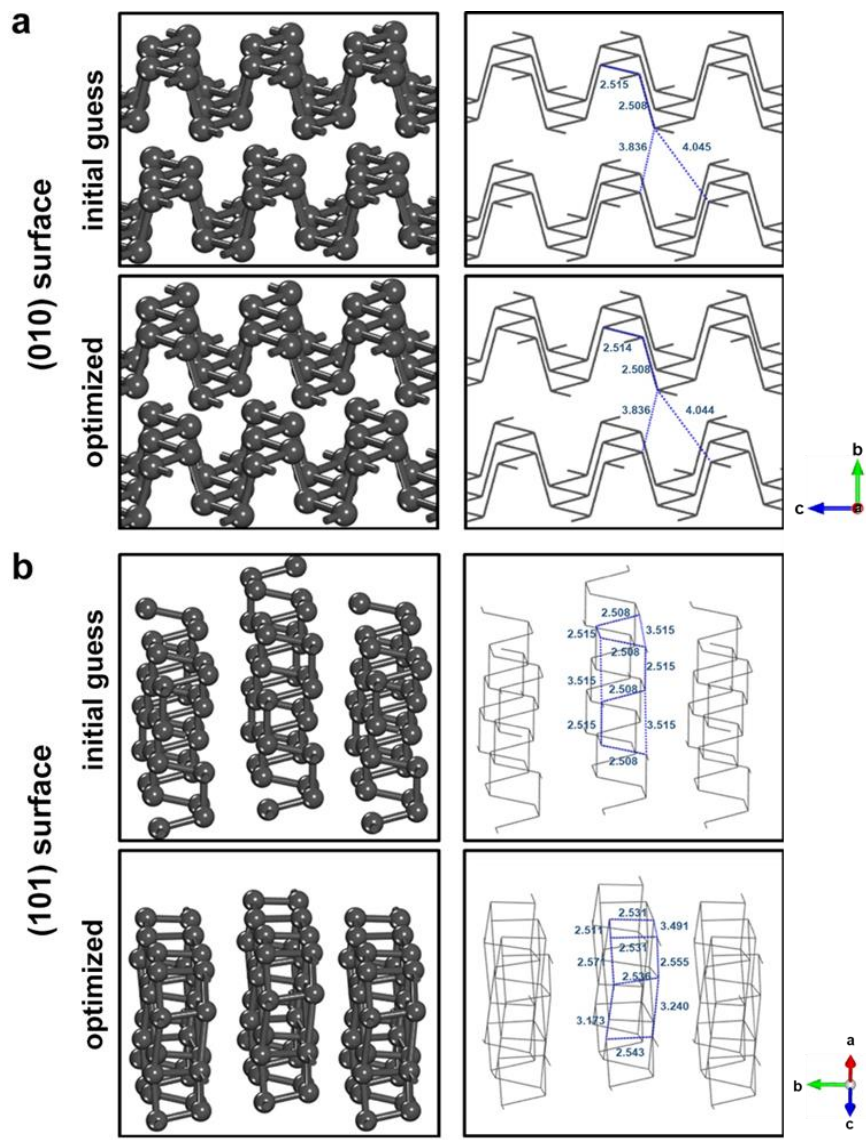

**Figure S1:** (a) Initial (top) and DFT-optimized (bottom) structures of the (010) terminated surface of b-As. The (010) surface, upon optimization, remains relatively unchanged as can be seen from the marked changes in the bond distances shown in the stick figure (units Å) on the right between. (b) Initial (top) and DFT-optimized (bottom) structures of the (101) terminated surface of b-As. The (101) surface, upon optimization, undergoes significant reconstruction as can be seen from the marked changes in the bond distances shown in the stick figure on the right. As atoms are shown in dark grey.

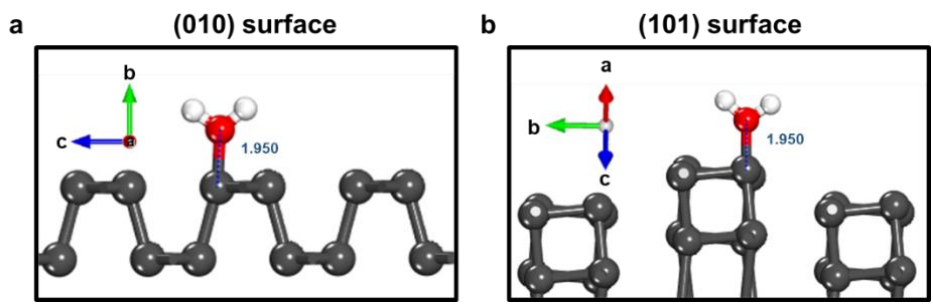

**Figure S2.** The initial structures of water adsorption over the bare (a) (010) and (b) (101) surfaces of b-As. In both cases the As-O distance was kept at 1.95 Å corresponding to low coverages of 1/8[th] of a surface on the (010) surface and 1/16[th] of a surface on the (101) surface. In these structures, As atoms are in dark grey, O in red, and H in white.

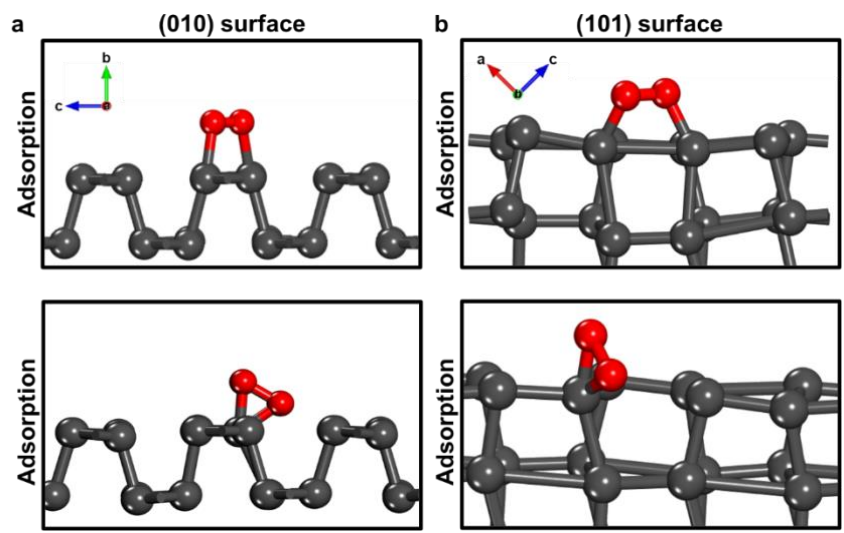

**Figure S3:** DFT-optimized structures for the other modes of O₂ adsorption: di-σ adsorption (top) and π-adsorption (bottom) on (a) (010) and (b) (101) b-As surface. In these structures, As atoms are in dark grey and O in red.

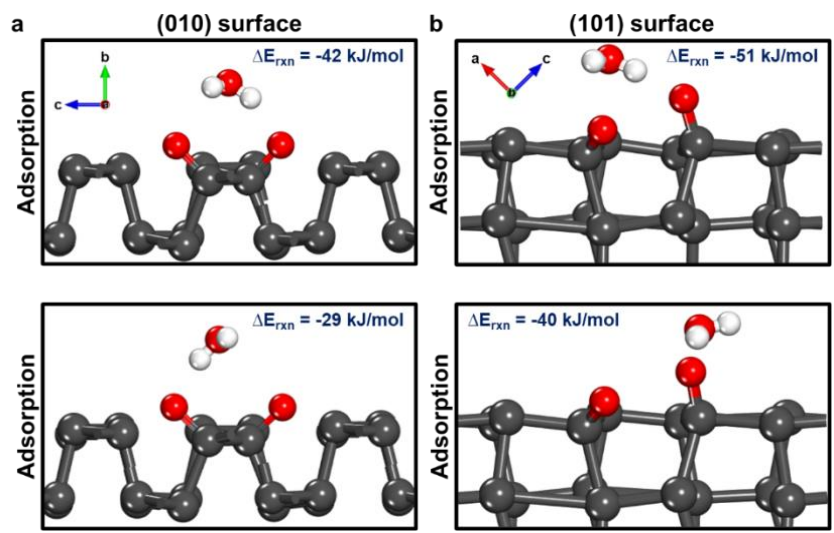

**Figure S4:** DFT-optimized conformational structures (top and bottom) for molecular adsorption of the water molecule over the two oxidized As atoms present on the (a) (010) and (b) (101) b-As surfaces. Low coverages of $1/4^{th}$ of a surface on the (010) surface and $1/8^{th}$ of a surface on the (101) surface were considered. In these structures, As atoms are in dark grey, O in red, and H in white.

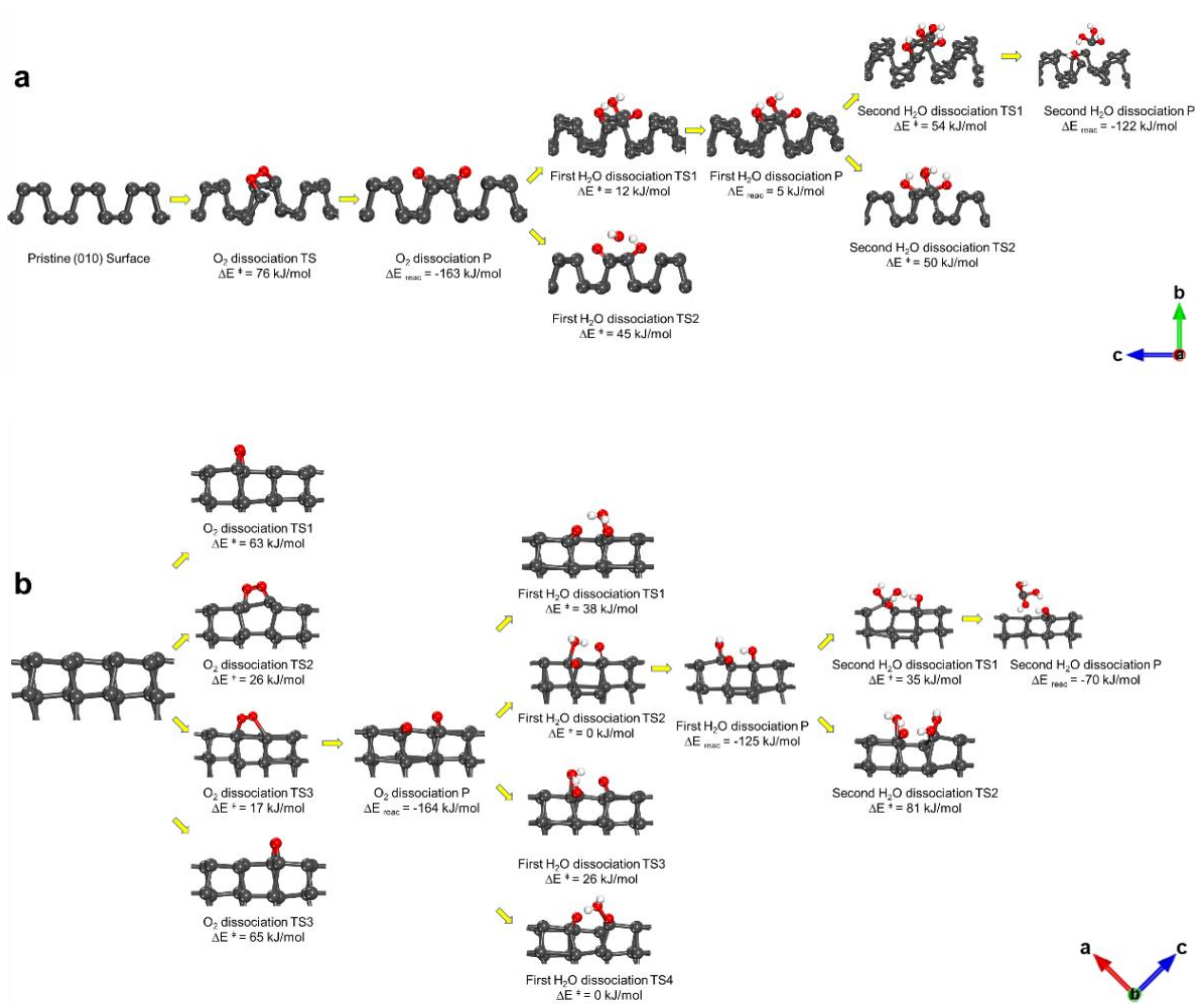

**Figure S5.** Flow diagram depicting the possible mechanistic routes involved in the sequential dissociation of oxygen followed by the dissociation of two water molecules with the corresponding reaction barriers and energies over the (a) (010) and (b) (101) b-As surface. In these structures, As atoms are in dark grey, O in red, and H in white.

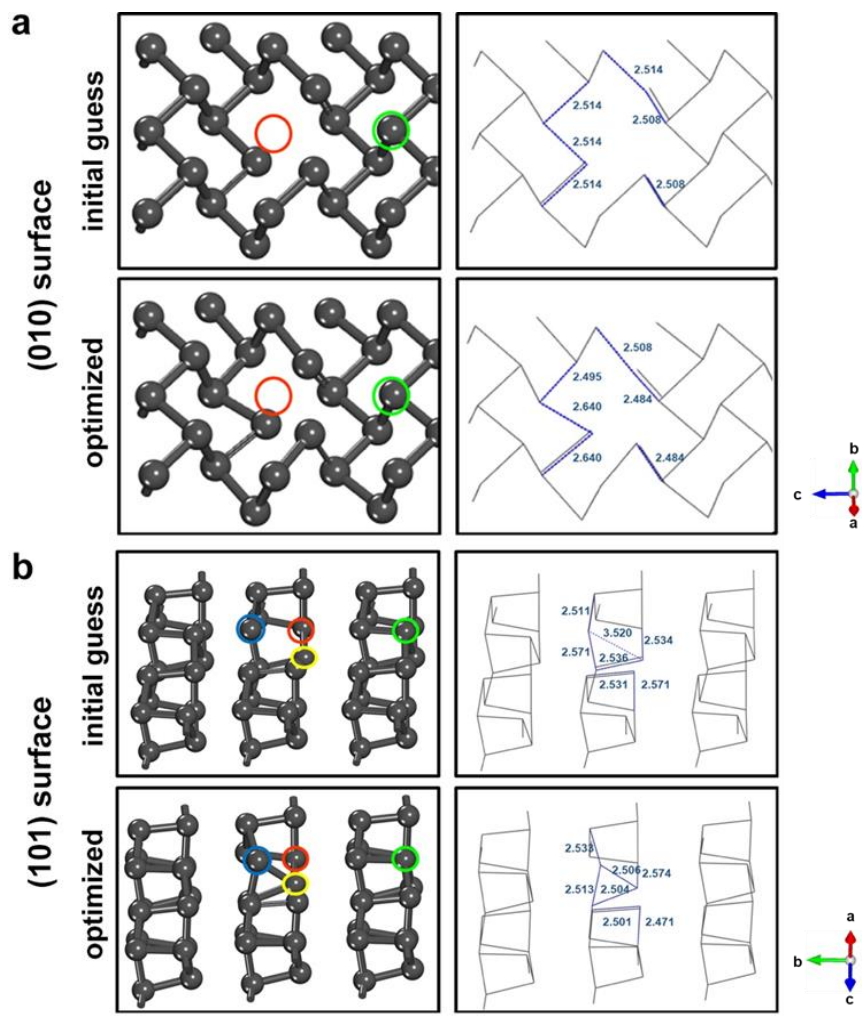

**Figure S6.** (a) Initial (top) and DFT-optimized (bottom) structures of the (010) b-As surface with an As vacancy (shown in the red circle). The green circle shows the As atom in a neighbouring site. The (010) surface of b-As with As vacancy remains relatively unchanged upon optimization with only minor changes in the bond distances (marked in the stick figure, in units Å). (b) Initial (top) and DFT-optimized (bottom) structures of the (101) b-As surface with As vacancy (shown in the red circle). The green circle shows a similar As atom in a neighbouring site. The (101) surface with the As vacancy reconstructs in such a way that the under-coordinated As atom (highlighted in blue) binds with another under-coordinated As atom present in the layer beneath it (highlighted in yellow). The bond distance optimizes from 3.520 to 2.506 Å.

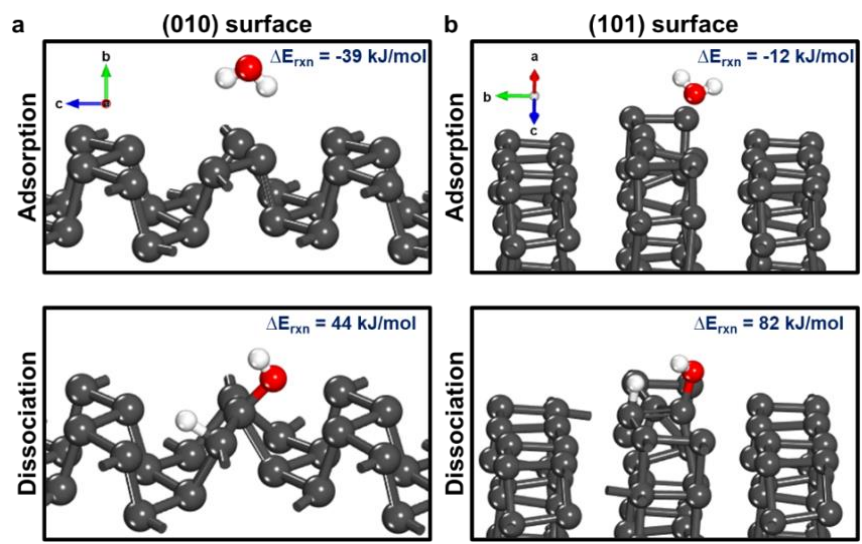

**Figure S7.** DFT-optimized structures for adsorption (top) and dissociation (bottom) of the H$_2$O molecule on (a) (010) and (b) (101) b-As surface with a As vacancy. In these structures, As atoms are in dark grey, O in red, and H in white.

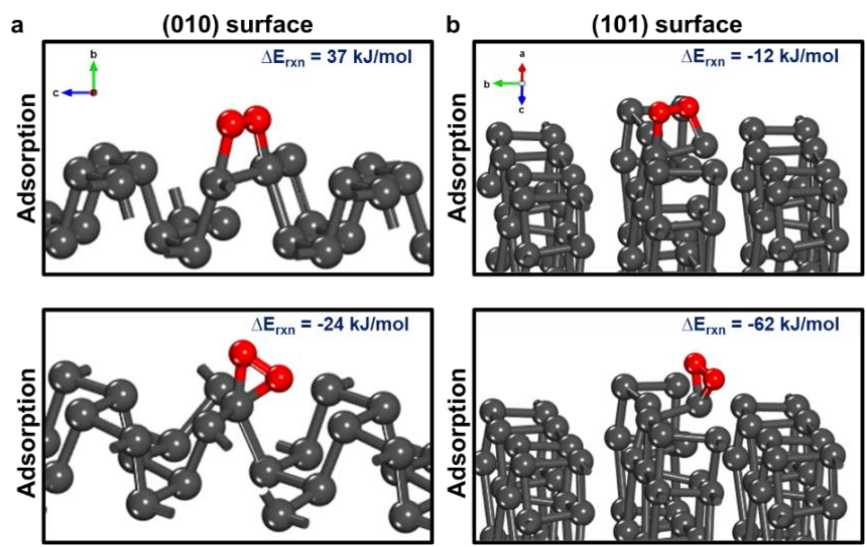

**Figure S8.** DFT-optimized structures for the other modes of O$_2$ adsorption: di-σ adsorption (top) and π-adsorption (bottom) on (a) (010) and (b) (101) b-As surface with a As vacancy. In these structures, As atoms are in dark grey and O in red.

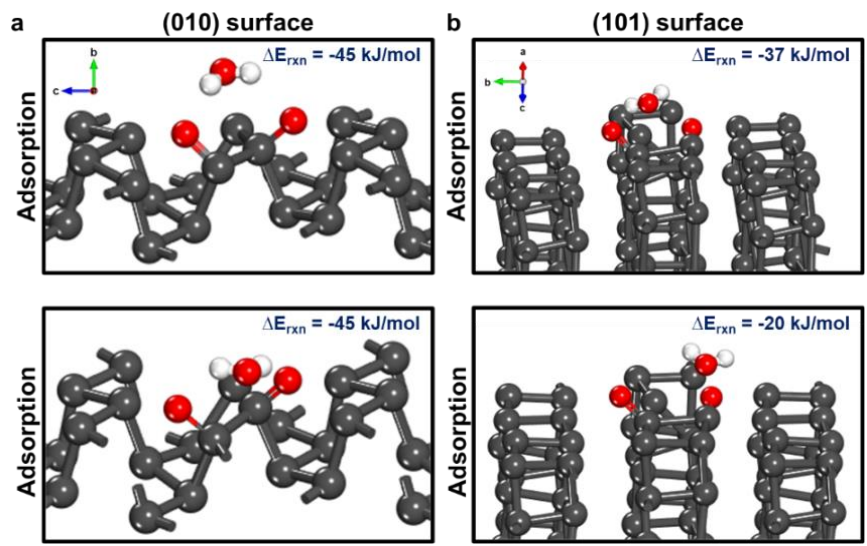

**Figure S9.** DFT-optimized two different conformational structures (top and bottom) for molecular adsorption of the water molecule over the two oxidized As atoms present on the (a) (010) and (b) (101) b-As surface with As vacancy. In these structures, As atoms are in dark grey, O in red, and H in white.

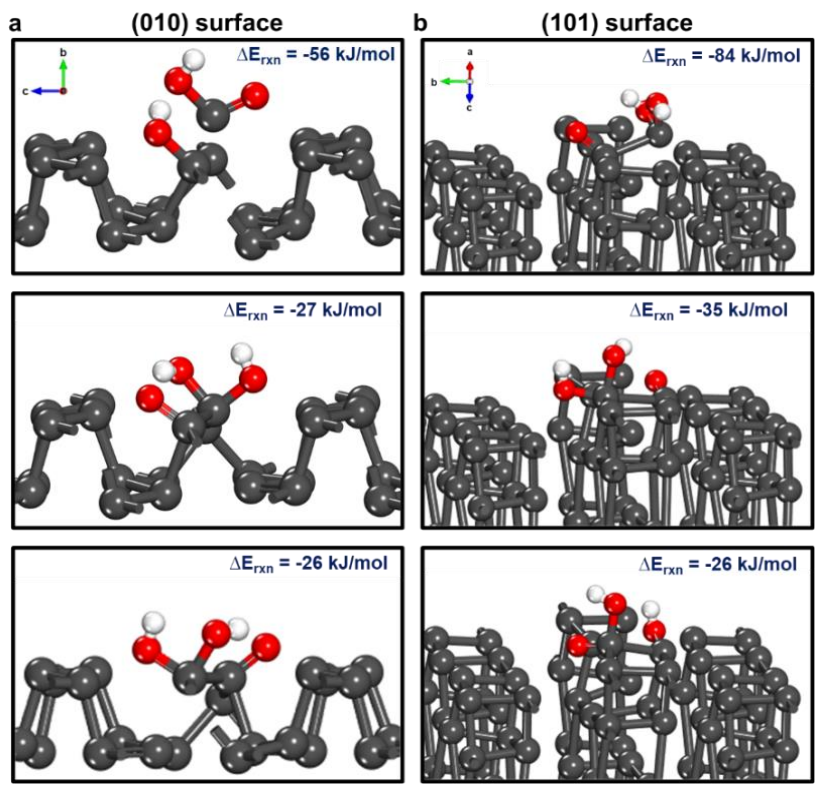

**Figure S10.** DFT-optimized structures of the other possible higher energy configurations for dissociation of the first water molecule over the oxidized (a) (010) and (b) (101) b-As surface with a As vacancy. In these structures, As atoms are in dark grey, O in red, and H in white.

**Supplementary discussion**

*Configurational sampling for the dissociation of oxygen and water over the oxidized As surfaces.*

While the surface sites on the DFT-optimized (010) black As (b-As) surface are identical, the surface sites on the DFT-optimized (101) surface are different. The differences arise due to the nature of bonding on the top layer as well as the layer below it. We identified four different pairs of vicinal sites on the (101) surface, which are either strongly covalently bonded on both the top and bottom layer or are strongly covalently bonded on either one. The sampling of the $O_2$ dissociation barriers was first carried out over these sites as shown in $O_2$ dissociation TS1-4 in Fig. S5.

Next, when water is dissociated over an oxidized b-As surface that contains dissociated oxygen atoms adsorbed on arsenic (As-O), it can undergo the following set of possible reactions.
As shown in the structure labelled "First $H_2O$ TS1" in Fig. S5, water can either be activated over a single As-O site with the dissociated OH entity adsorbing to the As site. The resulting H combines with the adsorbed oxygen (O) to form $As(OH)_2$:

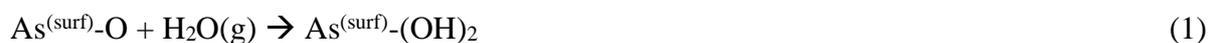

$$As^{(surf)}\text{-}O + H_2O(g) \rightarrow As^{(surf)}\text{-}(OH)_2 \qquad (1)$$

A second path for the dissociation of water involves the dissociation of a water molecule over two As-O sites, forming AsO(OH) and As(OH), as shown in the structure "First $H_2O$ TS2" in Fig. S5, is:

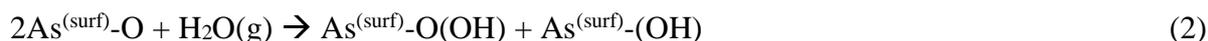

$$2As^{(surf)}\text{-}O + H_2O(g) \rightarrow As^{(surf)}\text{-}O(OH) + As^{(surf)}\text{-}(OH) \qquad (2)$$

Both routes of water dissociation over (010) and (101) terminated surfaces of b-As were considered in the analysis of the minimum energy barrier pathways as shown in Fig. S5.

While the dissociated hydrogen binds with the dissociated oxygen adsorbed over arsenic in the above two reactions, we also looked at the case where the hydrogen instead binds directly to the arsenic atom rather than with the oxygen atom. The dissociation of water via this mechanism, however, was either found to be highly endothermic or the resultant water dissociated structure did not optimize properly. The structure resulted in the hydrogen either bound to the oxygen or resulted in recombination of H and OH to reform water. These results suggest that the adsorption of hydrogen over an As atom is not favored, which is also consistent with the high endothermic energies (163 kJ/mol for (010) surfaces, 101 kJ/mol for (101) b-As surfaces) observed for water dissociation over b-As surfaces. Thus, for further analysis involving the second water dissociation step and the water dissociation over surfaces with As vacancy defect, this particular mode of dissociation was not considered.

The analysis for the second water dissociation step over the oxidized (010) and (101) b-As surfaces was similarly repeated using the configurations obtained from the first water molecule dissociation step to isolate the minimum energy barrier pathway for degradation of b-As. Starting from the final states obtained from the first water dissociation step, the following reactions were considered for the second water molecule dissociation step:

$$As^{(surf)}\text{-}(OH)_2 + As^{(surf)}\text{-}O + H_2O(g) \rightarrow As\text{-}(OH)_3(g) + As^{(surf)}\text{-}OH \qquad (3)$$

$$As^{(surf)}\text{-}O(OH) + As^{(surf)}\text{-}OH + H_2O(g) \rightarrow 2As^{(surf)}\text{-}(OH)_2 \qquad (4)$$

These are given as "Second $H_2O$ dissociation TS1" and "Second $H_2O$ dissociation TS2", respectively, in Fig. S5(a).

The possible mechanistic routes discussed above for both oxygen and water dissociation over the (010) and (101) b-As surfaces along with the relevant optimized structures, transition states and the reaction barriers and reaction energies are summarized in Fig. S5. The most favorable pathway among the various mechanistic routes examined was determined by considering the steps with the lowest energy barrier. Following these routes, it can be seen that the formation of As(OH)$_3$, which ultimately leads to the volatilization of the As atom from the surface, was found to be most favorable water dissociation product for the (101) b-As surface (as discussed in the main text).

For the defect surfaces examined comprised of As vacancies, a similar configurational sampling approach that examines the most exothermic structures was used to sample the different configurations for the first water dissociation which are shown in Fig. S10 with the minimum energy structures reported in Fig. 6 of the main text.